\documentclass[12pt,preprint]{aastex}

\begin{document}

\title{The Onset of Spiral Structure in the Universe}

\author{Debra Meloy Elmegreen\altaffilmark{1} and
Bruce G. Elmegreen\altaffilmark{2}}

\altaffiltext{1}{Vassar College, Dept. of Physics and Astronomy, Poughkeepsie, NY 12604}
\altaffiltext{2}{IBM Research Division, T.J. Watson Research Center, Yorktown Hts., NY 10598}

\begin{abstract}
The onset of spiral structure in galaxies appears to occur between redshifts 1.4
and 1.8 when disks have developed a cool stellar component, rotation dominates
over turbulent motions in the gas, and massive clumps become less frequent.
During the transition from clumpy to spiral disks, two unusual types of spirals
are found in the Hubble Ultra Deep Field that are massive, clumpy and irregular
like their predecessor clumpy disks, yet spiral-like or sheared like their
descendants. One type is ``woolly'' with massive clumpy arms all over the disk
and is brighter than other disk galaxies at the same redshift, while another
type has irregular multiple arms with high pitch angles, star formation knots
and no inner symmetry like today's multiple-arm galaxies. The common types of
spirals seen locally are also present in a redshift range around $z\sim1$,
namely grand design with two symmetric arms, multiple arm with symmetry in the
inner parts and several long, thin arms in the outer parts, and flocculent, with
short, irregular and patchy arms that are mostly from star formation. Normal
multiple arm galaxies are found only closer than $z\sim0.6$ in the UDF. Grand
design galaxies extend furthest to $z\sim1.8$, presumably because interactions
can drive a two-arm spiral in a disk that would otherwise have a more irregular
structure. The difference between these types is understandable in terms of the
usual stability parameters for gas and stars, and the ratio of the velocity
dispersion to rotation speed.
\end{abstract} 
\section{Introduction}

Galaxies at redshifts beyond $z\sim1$ become increasing irregular
\citep{abraham96, conselice05, lotz06, e07, cameron11, shapley11, buitrago13,lee} as a
result of rapid gas accretion \citep{queyrel12}, strong gravitational
instabilities \citep{bournaud07,genzel08,ceverino10,wisnioski11}, and distortions
produced by massive clumps \citep{ee05} and mergers
\citep{kartaltepe12,puech12,kaviraj13,mclure13}.  Most of this change is driven by
increasing self-gravity in the gaseous component of the disk, which increases in
relative mass and Mach number toward higher redshifts
\citep{tacconi13,bothwell13}. Local disks like the Milky Way have a relatively
small gas fraction in the inner few scale lengths, and self-gravity operates
somewhat separately in the stars and gas, making spiral arms in the stars on large
scales, and molecular clouds in the gas on small scales.

Star formation processes may vary also with these morphologies, from what is
presumably a collapse of gas to dense cores as a result of the strong
instabilities at high redshift, to secondary instabilities inside gas that is
first compressed by quasi-stable spirals or previous generations of star
formation. The star formation efficiency and maximum mass of a bound stellar
cluster could change too, perhaps explaining the prevalence of massive globular
clusters in early phase disks \citep{shap10,elmegreen12}.

A number of recent surveys have extended studies of galaxy structure and
evolution through much larger fields and at longer wavelengths than the original
HUDF optical study, including COSMOS (Cosmic Evolution Survey; \cite{scoville})
and CANDELS (Cosmic Assembly Near-Infrared Extragalactic Legacy Survey;
\cite{grogin}). These broad surveys provide statistical samples of galaxies and
are useful for studying global galaxy properties over cosmic timescales. One of
the largest recent studies includes nearly 1700 galaxies from CANDELS over the
redshift range $z=1.4-2.5$ to compare star formation properties and morphology
based on $H$-band light profiles and other global indicators such as the Gini
coefficient, concentration, and asymmetry \citep{lee}. They conclude that the
basic Hubble sequence of disk and spheroidal galaxies was set by $z\sim$2.

We are interested in understanding the physical processes in the intermediate
states of an evolving galaxy disk, between the high-redshift clumpy phase and the
low-redshift spiral phase. Do the clumps gradually shrink as the turbulence
dissipates and the stellar disk builds up \citep{cacciato12,elmegreen13}, or is
there a new morphology with massive substructures that are both spiral-like and
clump-like at intermediate redshifts?

In an unpublished study of $\sim 800$ spiral galaxies out to $z\sim 0.8$ in the
COSMOS fields based on WFPC2 optical images, we found a mixture of grand design,
multiple arm, and flocculent spirals over the whole redshift range, although the
fraction of multiple arm galaxies increased at lower redshift. In a study of
$\sim 200$ galaxies out to $z\sim1.4$ \citep{elmegreen09a} in the GEMS (Galaxy
Evolution from Morphology and SEDS; \cite{rix04} and Great Observatories Origins
Deep Survey (GOODS \cite{giavalisco04}) fields, we found these three
morphologies out to $z\sim1$ and also found two clumpy types
that overlapped in the same redshift range. One type had massive clumps with no evident
underlying disk, and the other type had equally massive clumps with an
underlying red disk. The ages and surface densities of the clumps and interclump
regions suggested that the former evolved into the latter as the disks grew from
dispersed clumps.  Two similar types were also found at $z\sim2$ in the UDF
\citep{elmegreen09b} where about half of the clumpy galaxies had red bulges that
were older and more massive than the other clumps, and the other half had no
bulge-like clumps. The ages of these clumps also suggested an evolutionary
sequence in the characteristic morphology of spirals, although at $z\sim2$ we
could not measure the faint interclump population.

Here we focus on early-phase spirals rather than late-phase clumps. The
combination of depth and resolution needed to examine spiral structure at high
redshift is available only in the Hubble Ultra Deep Field \citep[UDF;][]{beck06}
optical ACS images, so we examine morphology primarily from these images. We
supplement these with the WFC3 $H$-band images of the UDF when available,
although the decreased resolution and longer wavelength sometimes miss spiral
details. This paper examines disk galaxies as a function of redshift out to
$z\sim4$ in the UDF, dividing the spiral types into the usual classifications of
grand design (two symmetric arms), multiple arm (several long arms with an
irregular distribution in the outer parts) and flocculent (small patches of star
formation throughout the disk). We also look for unusual morphologies that could
be transition cases between clumpy and spiral disks.

We find grand design spirals out to at least $z=1.8$ in the UDF. The highest
redshift grand design spiral so far reported is in the 3-arm galaxy BX442 at
$z=2.18$ \citep{law12}. At these high redshifts, the HST ACS optical bands show
only the restframe ultraviolet, where star formation dominates. We should still
be able to see spiral structure in the restframe UV, however. GALEX images show
spiral arms in local galaxies in restframe NUV and FUV, in the familiar
beads-on-a-string pattern of star formation; each string is an arm in the
visible bands. Thus we can look for spirals in the higher redshift galaxies by
the filamentary alignment of their star formation clumps, as well as look in the
$H$-band directly for spiral arms. We present our observations in Section
\ref{sect:obs}, an interpretation in Section \ref{sect:int}, and our conclusions
in Section \ref{sect:conc}.

\section{Observations}
\label{sect:obs}

\subsection{Five Spiral Morphologies}

We examined by eye the morphological details of the spiral galaxies in the UDF
based on our catalog of 269 spirals  larger than 10 pixels in diameter
\citep{elmegreen05}. Of these, 184 are edge-on or too inclined to discern clear
spiral structure, 8 are disturbed tidal interactions or mergers, 5 should
probably have been classified as clump clusters, 20 have featureless disks
except for a bulge, and 11 have no redshifts. The remaining 41, which have
redshifts and clear spiral structure, are the ones included in the present
study. Both the optical ACS images and the $H$-band images were used when
available; about half of the 41 galaxies fell outside of the $H$-band field.

The galaxies are listed in Table 1, with photometric redshifts from
\cite{rafelski09}. The redshifts, which include UV data, are accurate to $\sim
0.1(1+z)$; 36 of the 41 galaxies have only one probability peak P(z), which
makes their redshifts relatively unambiguous. Three of the remaining four have a
secondary probability peak at a redshift less than 0.1 different from the
primary redshift; UDF 501 with a primary redshift of 1.37 has a secondary
probability peak nearby at $z=1.12$; UDF 9018 with a primary at $z=1.19$ has a
secondary at $z=0.78$. The table also includes the restframe $B$-band absolute magnitude,
the effective radius, the exponential disk scale length, the Sersic index, and
the spiral arm type, to be discussed more below.

\begin{figure}\epsscale{.9}
\plotone{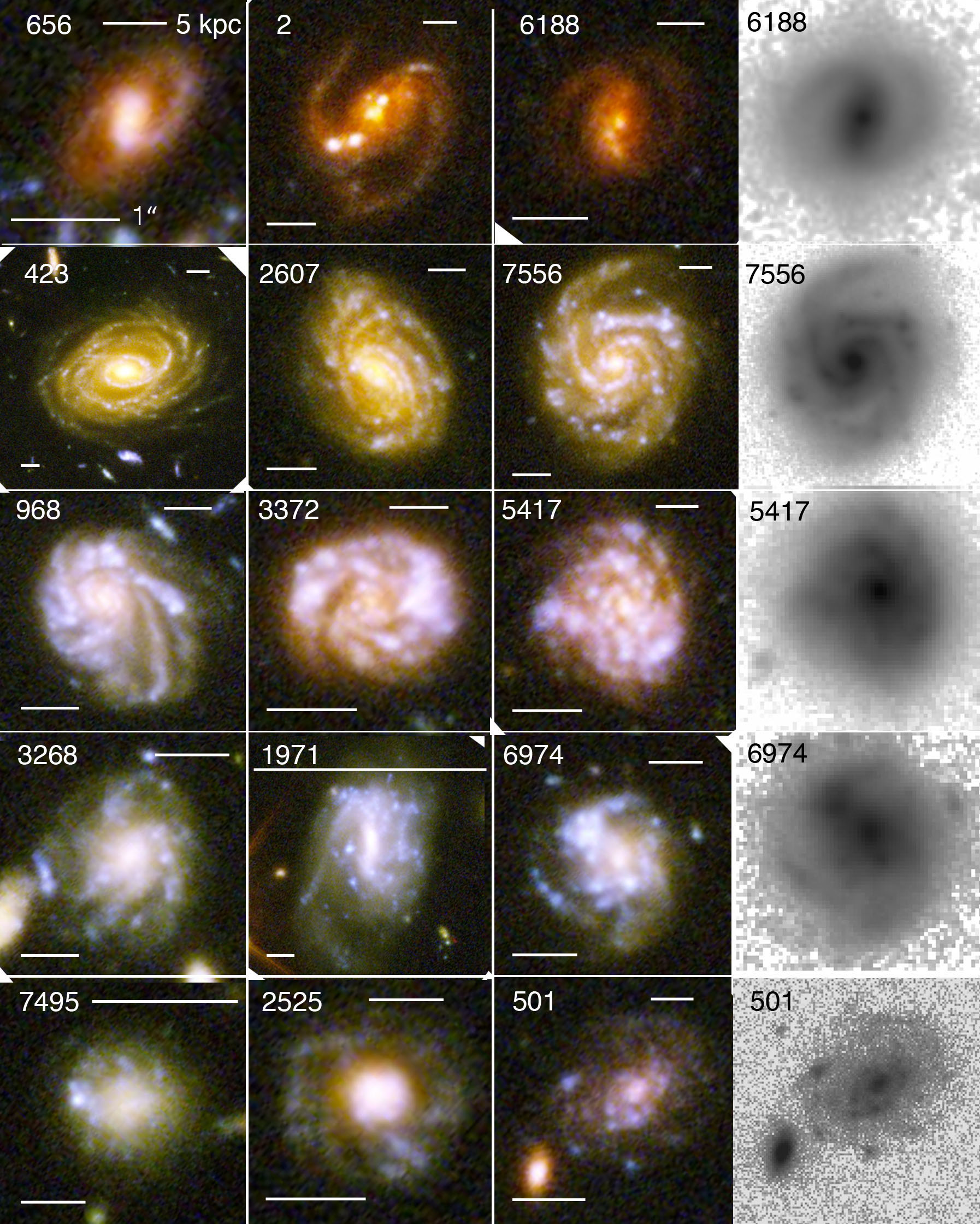}
\caption{Five spiral arm morphologies in the UDF with three examples of each
(in color) and an H-band image of the right-most example in black and white.
In each image, the bar on the lower left indicates an angular scale of 1'',
while the bar on the upper right indicates a linear scale of 5 kpc.
From top to bottom, the morphologies are: grand design, normal multiple arm, woolly,
irregular long-arm, and flocculent.  The properties of these galaxies are in Table 1.
What we call woolly consists of thick, patchy and long arms, in contrast to flocculent
galaxies which have numerous short and patchy arms. The irregular long-arm type consists
of numerous thin arms with bright beads of star formation in them and an overall
irregular or asymmetric structure.} \label{Fig1new}\end{figure}

Figure \ref{Fig1new} shows 15 spiral galaxies from the UDF in color images from
the UDF Skywalker that were made from ACS archival images in bands $B_{435}$,
$V_{606}$, and $i_{775}$ \citep{beck06}.  The black and white images are of the
third galaxy from the left in each row, taken in $H_{F160W}$ band with WFC3 for
HUDF09 and downloaded through the HST MAST HSLP archives \citep{bouwens}. The
UDF numbers from \cite{coe06} are indicated in the upper left of each frame.
White bars in the lower left of each frame indicate an angular scale of 1'',
while white bars in the upper right indicate a linear scale of 5 kpc. For
reference, the pixel scale of the optical HUDF images is 0.03'', with a FWHM
resolution of 0.1'' \citep{oesch}, while the $H$-band HUDF09 image has a pixel
scale of 0.06'' and a FWHM resolution of 0.16'' \citep{oesch10}.

Each row in Figure \ref{Fig1new} is a different morphological type. The top row
has three conventional grand design galaxies (designated type ``G'') with fairly
symmetric spiral arms, central bulges or bars, and smooth underlying disks. The
arms are open, wrapped at most $180^\circ$, and dotted with star formation knots
and streaks. Other grand design galaxies are in Table 1. The largest redshift
for this morphology in the UDF is $z= 1.8$ for UDF 9444.

In the second row are three normal-looking multiple arm galaxies (type ``M'')
that resemble a common type in the local universe. The arms in the outer parts
are long and thin, sometimes wrapping for $180^\circ$ or more, and the arms in
the inner parts are usually symmetric, sometimes with a short bar. The highest
redshift that we found for this type of spiral is 0.6 for UDF 2607 in the
figure. This is significantly smaller than the largest redshift for the grand
design spirals, of which there are four with $z>1$. Thin arms in multiple arm
galaxies like this require cool disks. It may be that this cool phase comes
later than the first grand design galaxies, whose structure may be induced by
strong bars or interactions even in hot disks (see below).

\begin{figure}\epsscale{1}
\plotone{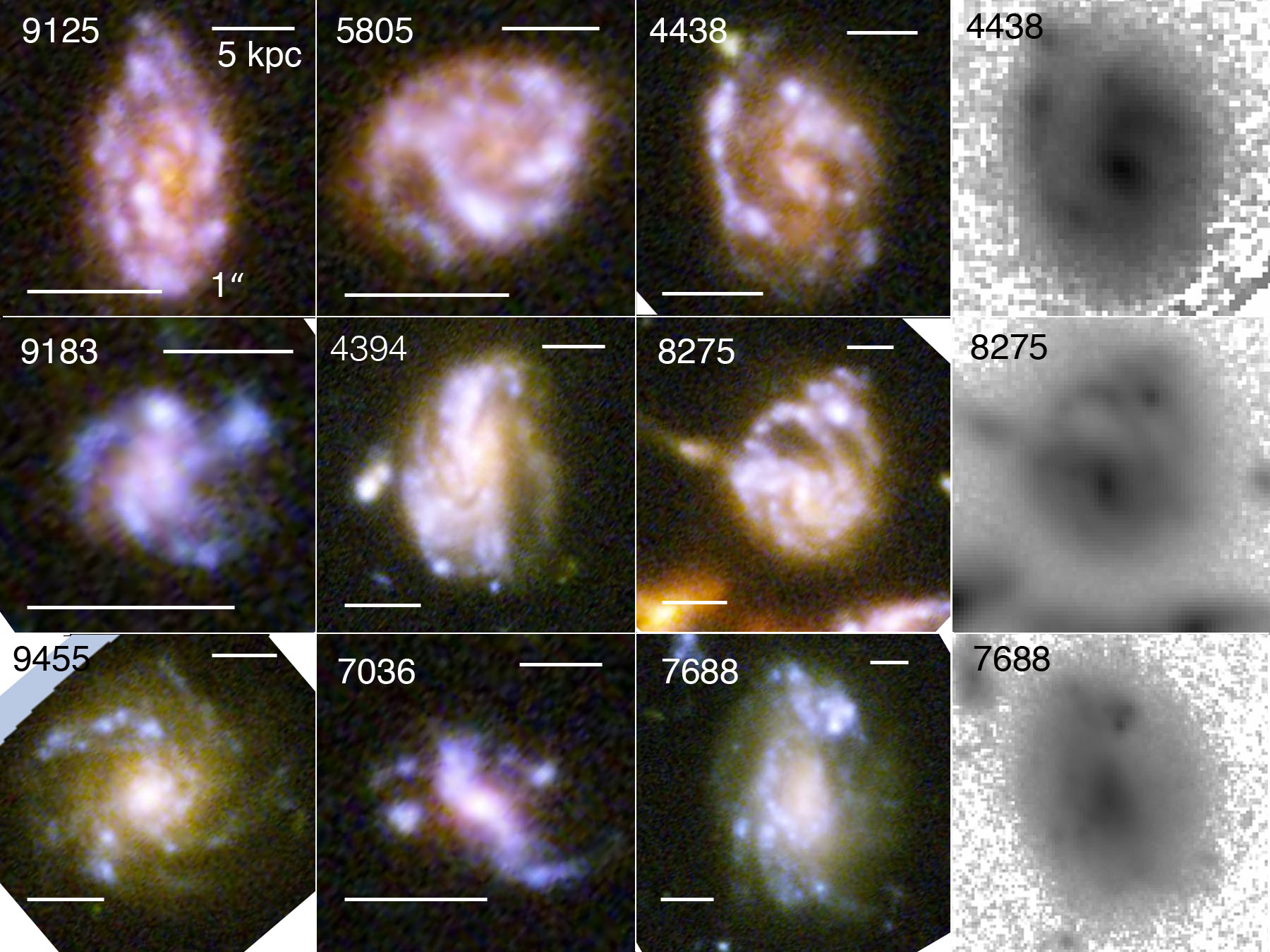}
\caption{Additional examples of the two new morphologies discussed here. The top
row contains more woolly galaxies and the bottom two rows contain more irregular
long-arm galaxies. Scales are for 1'' and 5 kpc, as in Figure \ref{Fig1new}.}\label{Fig2}\end{figure}

The third and fourth rows contain galaxy morphologies that are relatively
uncommon in the local universe. In the third row are multiple arm galaxies with
long, thick and bright arms that cover most of the disks.  The arms are clumpy
and the clumps are relatively large.  There are bright interarm clumps too.
Galaxies 968 and 5417 have offset nuclei. Local analogs also including a high
proportion with offset nuclei are types Ic in the Kiso Ultraviolet Survey of
Galaxies \citep{kiso-all}, whose clumps were found to be comparable in mass to
the clumps in UDF clumpy galaxies (not shown in the figures here), and which
also have multiple spiral arms \citep{elmegreen13}. The row-three galaxies in
Figure 1 differ from the Kiso type Ic galaxies, however, because the row-three
galaxies have arms that are relatively thick, bright and nearly continuously
filled with clumps.  We consider this a new morphology for spiral galaxies that
should be placed between purely clumpy galaxies at high redshift and
density-wave spiral galaxies at low redshift.  Their distinctive feature is
their thick patchy long arms. A term that comes to mind is ``woolly'' (type
``W'' in what follows) which is somewhat like ``flocculent'' but connotes
something more dense in structure. Three more of these galaxies are shown in the
top of Figure \ref{Fig2}.

The fourth row in Figure \ref{Fig1new} shows another odd type of structure
consisting of long, generally thin and multiple arms with haphazard
distributions and varying pitch angles. There are bright star-formation knots
along the arms, which is a common feature of most spiral arms, but no symmetry.
Some have small bars which seem unrelated to the spirals.  This is another
morphology that one might expect in an era of strong gravitational
instabilities, which readily produce large, shearing and chaotic-looking
material arms, but the clump size has decreased by now, perhaps because of a
reduced turbulent speed in the interstellar gas. Six more examples of this class
are in Figure \ref{Fig2}. We refer to this type descriptively again, calling
them ``irregular long'' arm galaxies (type ``IL''). That is, they are not like
classical irregulars which tend to have no spiral arms, nor are they like grand
design or multiple arm galaxies, which tend to have at least a little symmetry,
especially in the center.  To distinguish this class from those in the second
row, we sometimes call the latter ``normal'' multiple arm galaxies in what
follows.

The fifth row in Figure \ref{Fig1new} shows common looking flocculent galaxies
(type ``F'') with small patches of star formation and faint traces of underlying
spiral waves in the stars. Often the pitch angles are high, making the arms look
nearly radial (UDF 2525 in the figure, or UDF 9204, not shown). In the presence
of shear, such high pitch angles correspond to young ages, and that youth is
consistent with the patches being mostly star formation regions. In a flocculent
galaxy, most of the structure is from star formation and presumably, therefore,
instabilities in the gas.

To consider what effects might be caused by resolution limitations at different
redshifts, we note that in a $\Lambda$CDM cosmology \citep{spergel}, the angular
size of a 10 kpc disk is a constant $\sim$1.4'' between redshifts 0.6 and 4,
which applies to 15 of the 24 galaxies in Figures \ref{Fig1new} and \ref{Fig2}.
A glance at Table 1 reveals that there are galaxies of all 5 spiral types across
this redshift range, which implies that angular resolution is not systematically
varying among the different types. For example, UDF 2607 and 7556, two normal
multiple arm galaxies in Figure \ref{Fig1new}, have redshifts of 0.60 and 0.53,
respectively; UDF 968 and 3372, two woollies, have the same redshifts of 0.50
and 0.78. UDF 2525, a flocculent galaxy, has a redshift of 0.63, and UDF 9253
(not shown) is a grand design galaxy at a redshift of 0.52. Thus, their
different morphologies are not an artifact of different resolutions or different
restframe wavelengths. Further discussions of resolution and bandshifting
effects are described in \cite{elmegreen09a} regarding GEMS and GOODS
morphology.

Following the method in \cite{elmegreen09a}, we Gaussian-blur two of the present
galaxies (``M'' types UDF 423 at $z=0.29$ and UDF 7556 at $z=0.53$) to the
resolution of a third (``W'' type UDF 5417 at $z=1.11$), and also re-pixelate
the closest of these (UDF 423) to approximately match the pixel scales of the
other two. We start with the $z_{850}$ image of UDF 5417, the $V_{606}$ image of
UDF 7556, and the $B_{435}$ image of UDF 423; these correspond to similar
restframe wavelengths of $\sim$ 400-340 nm. The pixel scales are the same for
each pass band and the FWHMs of stars are similar: 3.12, 2.88, and 3.25,
respectively. At the distances to these galaxies, 1 px corresponds to 247 pc,
188 pc, and 129 pc, respectively. Thus the FWHM in UDF 5417 corresponds to 770
pc, which is 4.1 px in UDF 7556 and 5.9 px in UDF 423. We match the scale of the
most distant galaxy by blurring the closer galaxies by an amount equal to
$[({\rm px\; in }\;770{\rm pc})^2-({\rm px\; in\; FWHM})^2]^{1/2}$ and dividing
that by $(8 \ln 2)^{1/2}$ to get the Gaussian dispersion. These give Gaussian
blur dispersions of 1.22 px for UDF 7556 and 2.12 px for UDF 423. For the latter
we also block-average $2\times2$ pixels to approximate the linear resolution of
the distant galaxy, UDF 5417.

\begin{figure}\epsscale{1}
\plotone{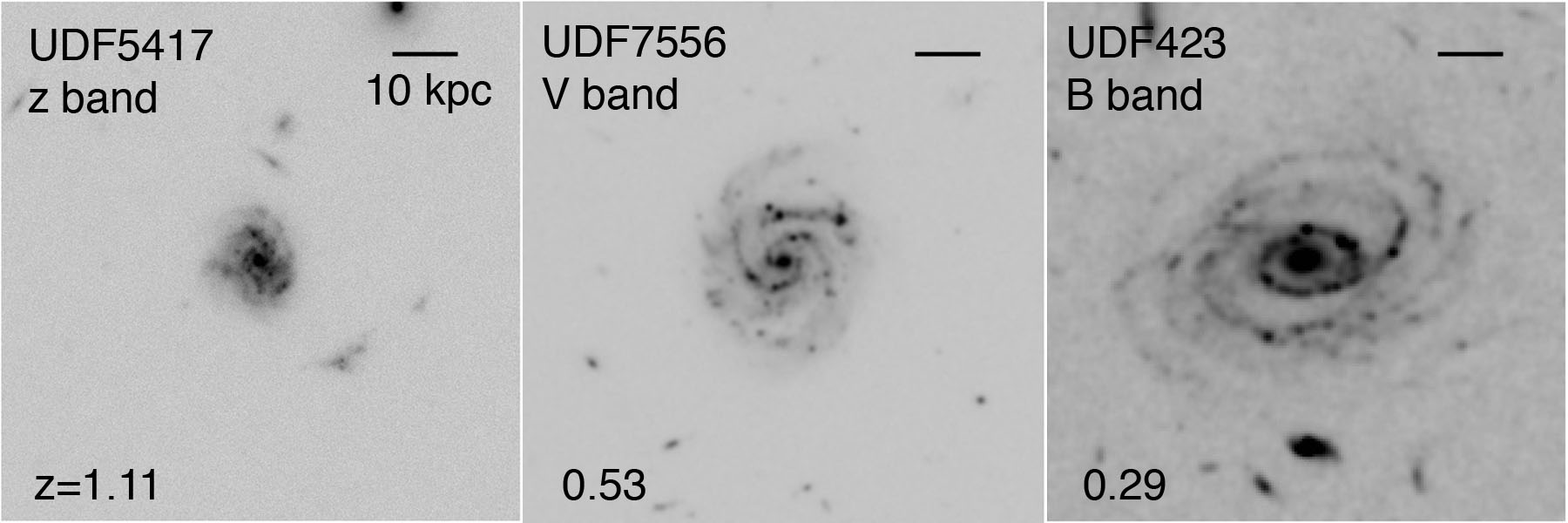}
\caption{UDF 5417 is a woolly galaxy; the other two are multiple arm,
and have been Gauss-blurred to match the resolution of the woolly. The different
bands all correspond to restframe $\sim$400 nm. The black line indicates a linear
scale of 10 kpc.}\label{Fig3}\end{figure}

The results of the Gaussian blurs are shown in Figure \ref{Fig3}, where each
galaxy is presented with approximately the same linear scale, pixel scale, and
restframe wavelength, as mentioned above. Evidently, the overall appearances of
the nearby multiple-arm galaxies (UDF 7556 and UDF 423) are not substantially
changed when the galaxies are blurred to look like the distant, woolly spiral
(UDF 5417). Bandshifting and resolution effects are not significant. The woolly
galaxy has bright irregular arms and large interarm star formation patches,
giving it a high surface brightness. This surface brightness is larger still,
when compared to the other two galaxies, if we consider that we have not
corrected for cosmological surface brightness dimming in Figure 3: that would
make the blurred galaxies dimmer in proportion to $(1+z)^{-4}$, which means a
factor of 0.28 and 0.14 for UDF 7556 and UDF 423, respectively. Such a reduction
in surface brightness would make the two multiple arm galaxies barely visible in
this figure if we were to include it.

\subsection{Properties of the Five Morphologies}

\begin{figure}\epsscale{1}
\plotone{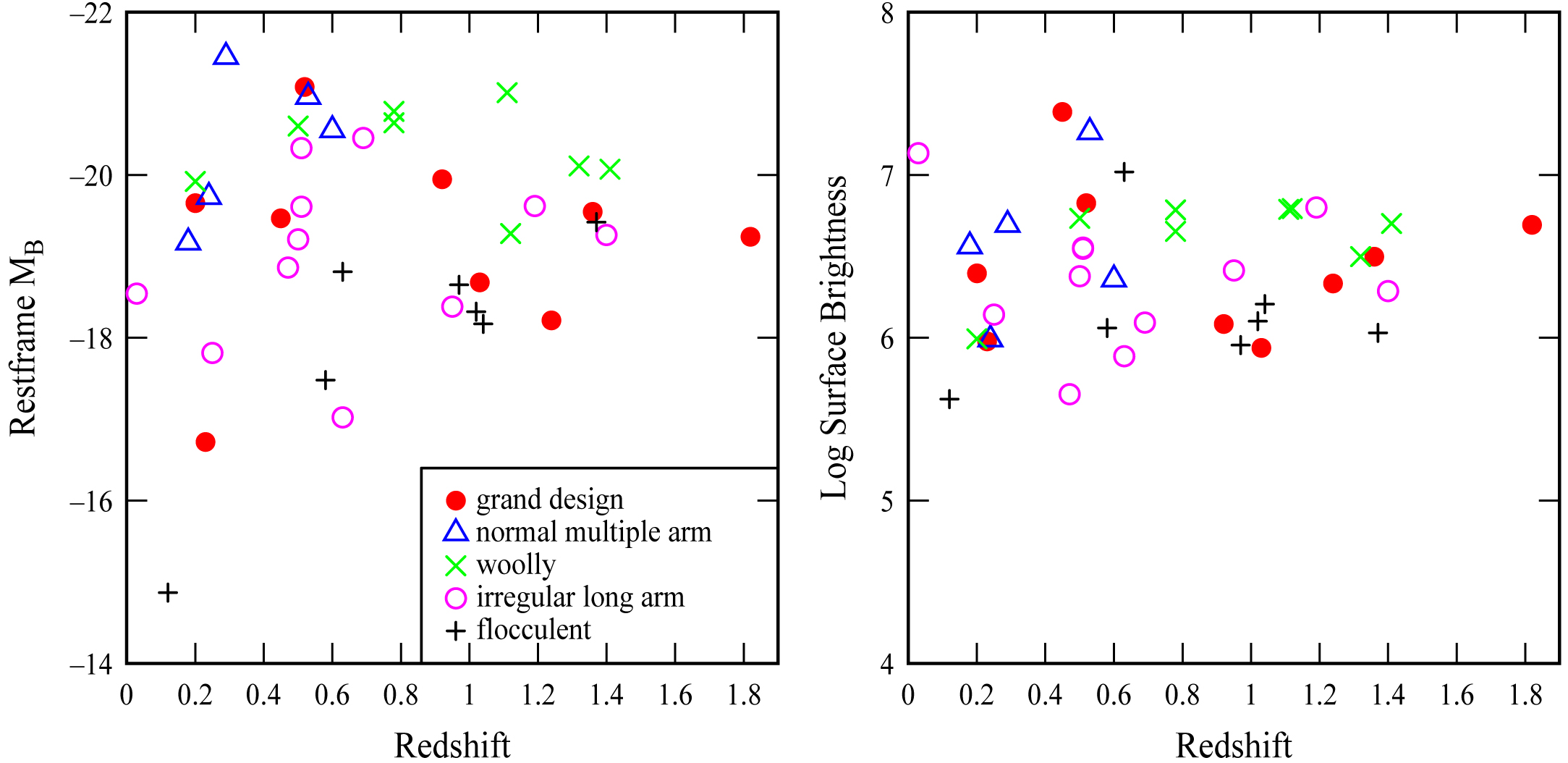}
\caption{The restframe $B$-band absolute magnitude of each galaxy in Table 1 is shown as
a function of the redshift in the left-hand panel.  The relative surface brightness,
$10^{-0.4M_B}/R_{\rm eff}^2$, is shown on the right (with $R_{\rm eff}$ in kpc).
Normal multiple arm galaxies and woolly galaxies are systematically larger and brighter
than the others, with higher surface brightnesses too. Flocculent galaxies are the
faintest. }\label{Fig4}\end{figure}

Figure \ref{Fig4} shows the redshift dependence of the restframe $B$-band absolute
magnitudes of the UDF spiral galaxies in Figures 1 and 2 and Table 1. It also
shows the surface brightnesses measured on a relative scale as
$10^{-0.4M_B}/R_{\rm eff}^2$ for $R_{\rm eff}$ the effective radius in Table 1,
in kpc.  The restframe $B$ absolute magnitudes come from COMBO17 (Classifying
Objects by Medium-Band Observations, a spectrophotometric 17-filter survey by
Wolf et al. 2003), but have been modified using an interpolation between $U$,
$B$, and $V$ bands to account for our use of \cite{rafelski09} redshifts instead
of COMBO17 redshifts. The effective radii come from GALFIT in a previous paper
based on restframe $B$ band \citep{e07} and were also adjusted to the
\cite{rafelski09} redshifts since the measurements originally used redshifts
from \cite{coe06}.

The overall brightness of our sample in absolute magnitudes increases with
redshift, partly as a selection effect for identifying these objects and partly
following the general increase in the universal star formation rate.  The
surface brightness is more constant, suggesting an increase in $R_{\it eff}$ with
restframe luminosity.  The woolly UDF spirals (green x-marks) and those with
normal multiple arms (blue triangles) are intrinsically brighter than the
others, on average, for all redshifts. The flocculents are fainter (as they are
locally). All types but the normal multiple arm galaxies span the full range of
redshifts, out to $z\sim1.4$ in most cases, and $z=1.8$ for one grand design
galaxy. We find no spiral galaxies of any type beyond this redshift, even
indirectly in the form of a spiral-like alignment of star formation clumps, nor
does the $H$-band image reveal spirals unnoticed in optical bands.

We measured the three largest star-forming clumps in several woolly and normal
multiple arm galaxies. For the normal multiple-arm galaxies, the three largest
clumps contribute 0.5\% to 2\% of the flux in the observer $i_{\rm 775}$ band
and 3\% to 5\% of the flux in the $B_{\rm 435}$ band. For the woolly galaxies,
the three largest clumps contribute 5\% to 15\% of the $i_{\rm 775}$ band flux,
and 10\% to 25\% of the $B_{\rm 435}$ band flux. Thus the largest clumps are
$\sim10$ times more luminous in the woolly galaxies than in the normal
multiple-arm galaxies.

\begin{figure}\epsscale{1}
\plotone{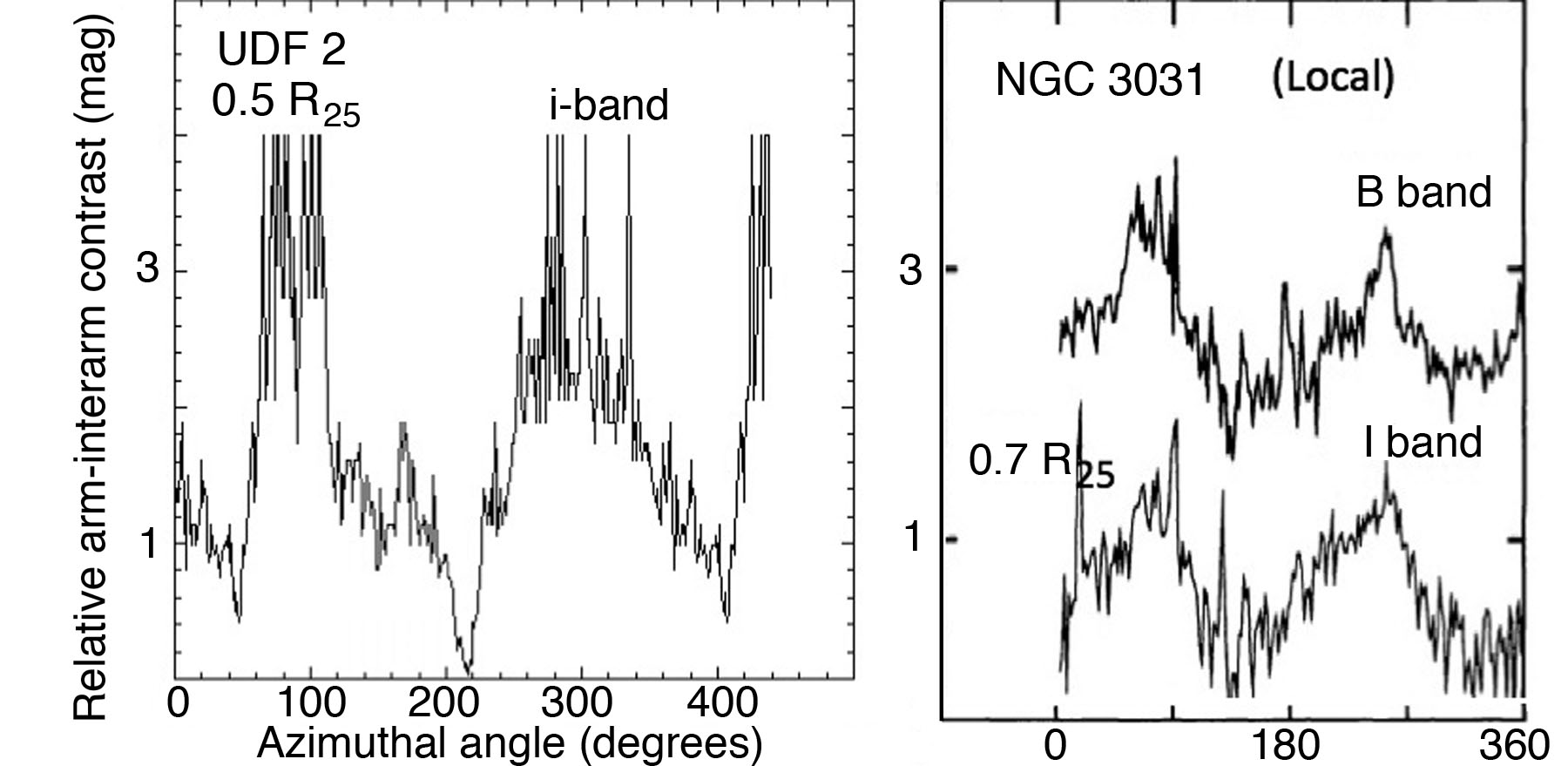}
\caption{Arm-interarm contrasts are shown for the grand design galaxies UDF 2 and
local NGC 3031.}\label{Fig5}\end{figure}

We also compared the arm-interarm contrasts in the UDF grand design galaxies
with those in local grand design galaxies, as measured from azimuthal profiles.
UDF 2 is shown along with local grand design spiral NGC 3031 \citep{elmegreen84}
in Figure \ref{Fig5}. The arm-interarm contrasts are about same in the UDF as in
the local grand designs, with the arms 1 to 2 mag arcsec$^{-2}$ brighter than
the interarms in the restframe $V$ band.

\begin{figure}\epsscale{.6}
\plotone{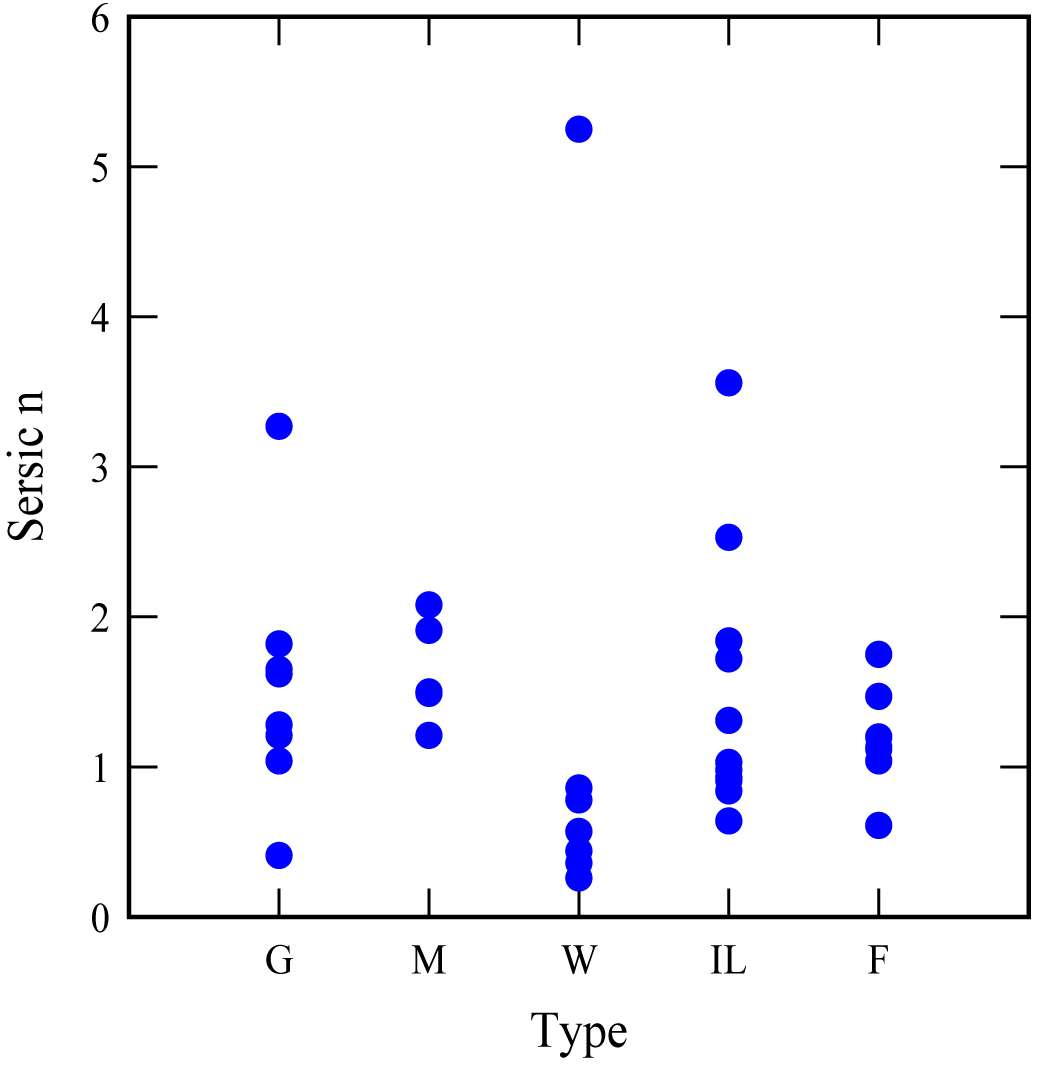}
\caption{Sersic index $n$ as a function of spiral morphological type. The
woolly spirals have systematically low $n$, consistent with their
having big clumps spread throughout the disk to give flatter light profiles. The
outliers with high $n$ have large central bulges.
}\label{Fig6}\end{figure}

The Sersic indices $n$ from \cite{e07} GALFIT measurements of the restframe $B$
images are shown in Figure \ref{Fig6} as a function of spiral morphology.
Because the galaxies in the present study are all disk galaxies, the indices average
around 1 for exponential disks. The outliers with large $n$ in the figure have
large central bulges.  The woolly types stand out as having low $n$, which is
consistent with their distributed light consisting of bright arms
and interarm clumps throughout the disk. This low-$n$ value for woollies is similar to
what was found for the most clumpy galaxies in the UDF \citep{e07}, further
suggesting that the woollies are an intermediate type between clumpy galaxies at
high redshift and normal spiral arm galaxies at low redshift.

\section{Interpretation}
\label{sect:int}

The five spiral morphologies discussed in the previous section correspond to
theoretical expectations based on the variation of several key parameters, such
as $Q_{\rm gas}$, $Q_{\rm star}$, and $L_{\rm Jeans}/R_{\rm gal}$, which are the
Toomre $Q$ values for gas and stars, and the ratio of the gaseous Jeans length
to the galaxy size (which is proportional to the square of the ratio of the gas
velocity dispersion to the rotation speed). The presence of strong interactions
and bars is a fourth determinant of structure. In addition, recent observations
of gas kinematics have shed light on how disk structures evolve with redshift.
Molecular gas observations by \cite{tacconi10, tacconi13} revealed that massive
star-forming galaxies have progressively higher gas fractions at $z\sim 1.2$ and
$z\sim2.3$ compared with the present. H$\alpha$ studies by \cite{reddy} and
\cite{erb06,erb} show higher star formation rates in high redshift galaxies,
with more massive galaxies reaching their peak star formation first. Kinematic
studies by \cite{for09} and \cite{genzel11} showed that massive clumpy galaxies
at $z\sim 2$ have high velocity dispersions, and \cite{kassin07} and
\cite{wright09} showed that velocity dispersion decreases with decreasing
redshift. These gas fractions and kinematics are consistent with numerical
simulations \citep{bournaud07,bournaud09,bournaud13} showing more massive clumps
in the early Universe than at present. Although the observations do not yet
extend to the five different spiral types studied here, with these parameters in
mind, we can reconstruct or predict what should be happening in their disks.

First of all, grand design galaxies should have relatively low $Q_{\rm star}$ so
the stellar disk is only marginally stable and readily forms strong stellar
spiral arms following perturbations from bars and companions. Gravitational
instabilities in the gas, involving $Q_{\rm gas}$, are not particularly
important for the overall grand design structure, and star formation itself is
secondary, following independent gravitational instabilities in the
spiral-shocked gas and other compressed gas. Normal multiple-arm galaxies (the
second row in Fig. \ref{Fig1new}) should also have relatively low $Q_{\rm
star}$, although perhaps in some cases not as low as in a strong grand-design
galaxy, but the multiple arm types tend to lack strong symmetric forcing, i.e.,
strong bars and interactions. In both cases, stellar gravity is relatively
strong, reflecting a high ratio of disk mass to halo mass in the optical parts.

Flocculent galaxies (row 5 in Figure \ref{Fig1new}) should have relatively weak stellar
gravity in the disk so that stellar spiral arms tend to be absent or weak. This
weakness corresponds to a somewhat higher $Q_{\rm star}$ and azimuthal parameter
$J$ \citep{lau78} (or lower $X$; Julian \& Toomre 1966), and to a relatively small
disk-to-halo mass ratio.  A difference between grand design and flocculent
galaxies with respect to the disk-to-halo ratio was measured in \cite{elmegreen90}
using outer HI rotation curves. A hot stellar disk, or a relatively large mass in
the thick disk, can make a galaxy flocculent in the absence of strong bars or
companion perturbations. Flocculent galaxies tend to be fainter than grand design
galaxies \citep{elmegreen87}, consistent with the luminosity class introduced by
\cite{vdb60}. This faintness is also evident from Figure \ref{Fig3}.

The two new types, the ``woolly'' and ``irregular long'' arm galaxies, also fit
into this three-parameter scenario. The woollies should be strongly
gravitationally unstable in both the stars and the gas ($Q_{\rm star}$ and
$Q_{\rm gas}$ relatively small) so spirals readily form even without bars or
companion perturbations, and stars form nearly everywhere in the
spiral-compressed gas.  The clumps are large so $L_{\rm Jeans}/R_{\rm gal}$ is
large, presumably because the velocity dispersion is relatively high compared to
the rotation speed -- high meaning several tens of percent, almost like what is
observed in clumpy galaxies \citep{for09}.  There should also be a significant
mass of stars in the disk, however, unlike the pure clumpies, and perhaps a
thick disk or spheroid as well, causing the disk instabilities to make spirals
rather than clumps \citep{bournaud09}. Irregular long-arm galaxies should also
have strong disk instabilities ($Q_{\rm star}$ and $Q_{\rm gas}$ relatively
small) but now $L_{\rm Jeans}/R_{\rm gal}$ is small too (making the clumps
small), presumably because the gaseous velocity dispersion has decreased.

The structure of these intermediate spiral types resembles that in the later
stages of clumpy-disk simulations by \cite{bournaud13}. Galaxies 968 and 1971 in
Figure 1 were also shown in Figure 15 of that paper. According to the
simulations, the transition from clumpy structure to spiral structure takes
about 1 Gyr, and the irregular long-arm structure occurs at a time between 500
and 800 Myr after the galaxy forms.  The gas in the simulations has a high
velocity dispersion of $\sim40-50$ km s$^{-1}$ and is gravitational unstable to
star formation, with Q$\sim1$. At later stages, $z\sim1$ or less, massive clumps
become relatively rare, disks begin to settle down, and rotation begins to
dominate over turbulent motions in the gas \citep{genzel06,kassin07,for09}. Bars
appear at about the same time \citep{sheth12}.

Considering the basic parameters that determine galactic structure, the disk
morphologies that have been identified at intermediate to high redshifts might
be placed into the following sequence for the development of spiral structure in
galaxies over time. The first stage is when the disk is nearly pure clumps and
there is relatively little interclump emission from older stars
\citep{elmegreen09a,elmegreen09b}. This corresponds to a gas-dominant phase when
interstellar turbulent speeds are also high \citep{bournaud07}. The second stage
is clumpy but with an interclump stellar population \citep{elmegreen09a},
presumably resulting from the dissolution of clumps. A classical bulge and thick
disk form in these first two stages. In subsequent stages, there are enough hot
stars from the earlier stages (in a thick disk and spheroid) that gravitational
instabilities make spirals rather than clumps \citep{bournaud09}. At first,
there may still be a relatively high gas fraction and turbulent speed, making
the morphology we identify here as ``woolly,'' namely with long and thick arms
covered with giant star forming regions. Then, as the gas cools and additional
gas accretes more slowly, the star-forming clumps get smaller even if the disk
stays highly unstable, forming spirals all over in an irregular way. This is the
irregular long-arm phase highlighted here. Finally the stellar instabilities
calm down as the stellar disk warms up \citep{cacciato12}, and the gas cools as
it accretes at a slower rate without the irregular forcing that formerly drove
fast turbulence. Gaseous instabilities also become weaker at these later phases,
when the gas fraction drops, and this, in addition to a decreasing star
formation rate, also limits the turbulent speed. Then normal multiple arm and
flocculent galaxies appear. Bars and strong interactions should make grand
design spirals in the main disks of the latter stages, but not in the pure-clump
phases, which have too few disk stars and disk stars that are too hot to
organize into a coherent spiral pattern. Presumably, tidal arms in the outer
regions of any of these galaxy types can be made by strong enough interactions
\citep[e.g.][]{law12}.

\section{Conclusions}
\label{sect:conc}

Spiral structure appears in the Hubble Ultra Deep Field by $z\sim1.8$ and most
prominently below $z\sim1.4$.  The observations of different spiral types are
consistent with the interpretation that clumpy disks form first and then
transition to spirals as the accretion rate and gas velocity dispersion
decrease, and the growing population of old fast-moving stars begins to dominate
the disk mass. These trends are consistent with kinematic observations and
numerical simulations. Grand design structures appear somewhat earlier than the
other spiral types in our survey, perhaps because strong interactions can make
symmetric tidal arms with grand design structure in a disk that would otherwise
have no visible spirals.

We are grateful to the referee for helpful comments.

\clearpage

\begin{deluxetable}{lcccccc}
\tabletypesize{\scriptsize}\tablecolumns{7} \tablewidth{0pt} \tablecaption{Galaxy properties}
\tablehead{\colhead{UDF}&
\colhead{z\tablenotemark{a}} &
\colhead{$M_{\rm B}$ (mag)\tablenotemark{b}} &
\colhead{$R_{\rm eff}$ (kpc)\tablenotemark{c}} &
\colhead{Disk Scalelength (kpc)\tablenotemark{c}} &
\colhead{Sersic $n$\tablenotemark{c}} &
\colhead{type\tablenotemark{d}}}
\startdata
Figure 1 & & & & & & \\
  656&   1.24&  -18.21&    3.0&    1.4&    3.3&G\\
    2&   0.92&  -19.95&    8.9&    3.0&    1.6&G\\
 6188&   1.03&  -18.68&    5.9&    2.5&    1.8&G\\
  423&   0.29&  -21.45&    8.8&    3.4&    1.5&M\\
 2607&   0.60&  -20.55&    8.5&    3.2&    2.1&M\\
 7556&   0.53&  -20.96&    3.6&    2.1&    1.2&M\\
  968&   0.50&  -20.60&    5.7&    2.1&    0.9&W\\
 3372&   0.78&  -20.78&    5.8&    3.3&    0.3&W\\
 5417&   1.11&  -21.01&    6.4&    2.2&    0.8&W\\
 3268&   0.25&  -17.81&    3.1&    1.5&    1.3&IL\\
 1971&   0.03&  -18.54&    1.4&    0.6&    1.7&IL\\
 6974&   0.50&  -19.21&    4.5&    2.0&    0.6&IL\\
 7495&   0.12&  -14.87&    1.5&    1.1&    0.6&F\\
 2525&   0.63&  -18.81&    1.8&    1.4&    1.7&F\\
  501&   1.37&  -19.42&    7.4&    3.8&    1.5&F\\
Figure 2 & & & & & & \\
 9125&   1.32&  -20.11&    5.9&    2.9&    0.6&W\\
 5805&   1.41&  -20.07&    4.6&    1.6&    0.4&W\\
 4438&   0.78&  -20.64&    6.3&    6.8&    0.4&W\\
 9183&   0.95&  -18.38&    3.0&    1.9&    0.8&IL\\
 4394&   0.51&  -20.33&    6.2&    4.0&    0.9&IL\\
 8275&   0.69&  -20.45&   11.0&    3.1&    1.7&IL\\
 9455&   0.47&  -18.86&    8.8&    2.6&    2.5&IL\\
 7036&   1.40&  -19.26&    5.1&    2.4&    3.6&IL\\
 7688&   2.58&  -16.83&    9.7&    7.0&    1.0&IL\\
Others & & & & & & \\
 8261&   1.36&  -19.55&    4.6&    2.8&    1.2&G\\
 9444&   1.82&  -19.24&    3.2&    2.4&    1.3&G\\
 9253&   0.52&  -21.08&    6.4&    3.0&    1.6&G\\
 3492&   0.20&  -19.65&    5.4&    3.8&    1.0&G\\
 4929&   0.45&  -19.47&    1.6&    1.2&    1.6&G\\
 6082&   0.23&  -16.72&    2.3&    0.8&    0.4&G\\
  328&   0.24&  -19.73&    8.9&    3.8&    1.9&M\\
 3822&   0.18&  -19.17&    3.6&    1.8&    1.5&M\\
 7112&   1.12&  -19.28&    2.9&    1.0&    0.4&W\\
 8049&   0.20&  -19.92&    9.7&    1.3&    5.2&W\\
 6862&   0.51&  -19.61&    4.4&    2.0&    1.8&IL\\
 6997&   0.63&  -17.02&    2.9&    2.0&    1.0&IL\\
 9018&   1.19&  -19.62&    3.3&    2.4&    0.9&IL\\
 1905&   0.58&  -17.48&    2.9&    1.5&    1.1&F\\
 5922&   1.02&  -18.32&    4.1&    3.4&    1.2&F\\
 9341&   1.04&  -18.17&    3.4&    2.3&    1.0&F\\
 9895&   0.97&  -18.65&    5.7&    3.6&    1.1&F
\enddata
\tablenotetext{a}{from Rafelski et al. (2009) except for UDF 328, which uses the
COMBO17 redshift} \tablenotetext{b}{from COMBO17, corrected to redshifts from
Rafelski et al. (2009)} \tablenotetext{c}{from Elmegreen et al. (2007),
corrected to redshifts from Rafelski et al. (2009)} \tablenotetext{d}{``G'' =
grand design (two prominent and symmetric arms), ``M'' = normal multiple arms,
usually symmetric and 2-arm in the inner parts, ``W'' = ``woolly'' in reference
to thick and patchy long arms, ``IL'' = ``irregular long'' arms, as distinct
from multiple arms because there is little symmetry, ``F'' = flocculent with
faint patchy and short arms.}
\end{deluxetable}

\clearpage

\clearpage

\clearpage

\clearpage

\clearpage

\clearpage

\clearpage


\begin{thebibliography}{}

\bibitem[Abraham et al.(1996)]{abraham96} Abraham, R., van den Bergh, S.,
    Glazebrook, K., Ellis, R., Santiago, B., Surma, P., \& Griffiths, R. 1996, ApJS, 107, 1

\bibitem[Beckwith et al.(2006)]{beck06} Beckwith, S.V.W., et al. 2006, AJ, 132,
1729

\bibitem[Bothwell et al.(2013)]{bothwell13} Bothwell, M. S., Smail, I.,
    Chapman, S.C., Genzel, R. et al. 2013, MNRAS, 429, 3027

\bibitem[Bournaud et al.(2007)]{bournaud07} Bournaud, F., Elmegreen, B.G.,  \&
    Elmegreen, D.M. 2007, ApJ, 670, 237

\bibitem[Bournaud \& Elmegreen(2009)]{bournaud09} Bournaud, F., \& Elmegreen, B.G.
    2009, ApJ, 694, L158

\bibitem[Bournaud et al.(2013)]{bournaud13} Bournaud, F., et al. 2013, ApJ, in press [arXiv:1307.7136]

\bibitem[Bouwens et al.(2011)]{bouwens} Bouwens, R.J. et al. 2011, ApJ, 737, 90S.

\bibitem[Buitrago et al.(2013)]{buitrago13} Buitrago, F., Trujillo, I.,
Conselice, C.J., \& H\"aussler, B. 2013, MNRAS, 428, 1460

\bibitem[Cacciato et al.(2012)]{cacciato12} Cacciato, M. Dekel,
    A., \& Genel, S. 2012, MNRAS, 421, 818

\bibitem[Cameron et al.(2011)]{cameron11} Cameron, E., Carollo, C. M., Oesch, P.
    A., Bouwens, R.J., Illingworth, G.D., Trenti, M., Labb\'e, I., \& Magee, D.
    2011, ApJ, 743, 146


\bibitem[Ceverino et al.(2010)]{ceverino10} Ceverino, D., Dekel, A.,
    Bournaud, F. 2010, MNRAS, 404, 2151

\bibitem[Coe et al.(2006)]{coe06} Coe, D., Ben\'itez, N., Sanchez, F., Jee, M.,
    Bouwens, R., \& Ford, H.  2006, AJ, 132, 926

\bibitem[Conselice et al.(2005)]{conselice05} Conselice, C. J., Blackburne,
    J. A., Papovich, C. 2005, ApJ, 620, 564

\bibitem[Elmegreen \& Elmegreen(1984)]{elmegreen84} Elmegreen, D.M., \& Elmegreen,
    B.G. 1984, ApJS, 54, 127

\bibitem[Elmegreen \& Elmegreen(1987)]{elmegreen87} Elmegreen, D.M., \& Elmegreen,
    B.G. 1987, ApJ, 314, 3

\bibitem[Elmegreen \& Elmegreen(1990)]{elmegreen90} Elmegreen, D.M. \& Elmegreen,
    B.G. 1990, ApJ, 364, 412

\bibitem[Elmegreen \& Elmegreen(2005)]{ee05} Elmegreen, B.G., \&
    Elmegreen, D.M. 2005, ApJ, 627, 632

\bibitem[Elmegreen et al.(2005)]{elmegreen05} Elmegreen, D.M., Elmegreen,
    B.G., Rubin, D.S., \& Schaffer, M. 2005, ApJ, 631, 85

\bibitem[Elmegreen et al.(2007)]{e07} Elmegreen, D.M., Elmegreen, B.
    G., Ravindranath, S., \& Coe, D.A. 2007, ApJ, 658, 763

\bibitem[Elmegreen et al.(2009a)]{elmegreen09a} Elmegreen, D.M., Elmegreen,
    B.G., Marcus, M., Shahinyan, K., Yau, M., \& Petersen, M. 2009a, ApJ, 701, 306

\bibitem[Elmegreen et al.(2009b)]{elmegreen09b} Elmegreen, D.M., Elmegreen,
    D.M., Fernandez, M.X., \& Lemonias, J.J. 2009b, ApJ, 692, 12

\bibitem[Elmegreen et al.(2012)]{elmegreen12} Elmegreen, B.G., Malhotra,
    S., \& Rhoads, J. 2012, ApJ, 757, 9

\bibitem[Elmegreen et al.(2013)]{elmegreen13} Elmegreen, B.G., Elmegreen, D.M.,
    S\'anchez Almeida, J., Mu\~noz-Tu\~n\'on, C., Dewberry, J., Putko, J.,
    Teich, Y., \& Popinchalk, M. 2013, ApJ, 2013, ApJ, 774, 86

\bibitem[Erb et al.(2006)]{erb06}Erb, D. et al. 2006, ApJ, 646, 107

\bibitem[Erb et al.(2009)]{erb}Erb, D. et al. 2006, ApJ, 647, 128

\bibitem[F\"orster Schreiber et al.(2009)]{for09} F\"orster Schreiber N. M.
    et al., 2009, ApJ, 706, 1364

\bibitem[Genzel et al.(2006)]{genzel06} Genzel R. et al., 2006, Nature, 442, 786

\bibitem[Genzel et al.(2008)]{genzel08} Genzel R. et al., 2008, ApJ, 687, 59

\bibitem[Genzel et al.(2011)]{genzel11} Genzel R. et al., 2011, ApJ, 733, 101

\bibitem[Giavalisco et al.(2004)]{giavalisco04} Giavalisco, M., Ferguson, H. C.,
    Koekemoer, A.M., Dickinson, M. et al. 2004, ApJ, 600, L93

\bibitem[Grogin et al.(2011)]{grogin}Grogin, N.A., et al. 2011, ApJS, 197, 35

\bibitem[Julian \& Toomre(1966)]{julian66} Julian, W. H., \& Toomre, A. 1966, ApJ,
    146, 810

\bibitem[Kartaltepe et al.(2012)]{kartaltepe12} Kartaltepe, J.S.,
    Dickinson, M., Alexander, D.M., Bell, E.F., et al. 2012, ApJ, 757, 23

\bibitem[Kassin et al.(2007)]{kassin07} Kassin, S.A., Weiner, B.J.,
    Faber, S.M., Koo, D.C., et al. 2007, ApJ, 660, L35

\bibitem[Kaviraj et al.(2013)]{kaviraj13} Kaviraj, S., Cohen, S., Windhorst,
    R.A., Silk, J., O'Connell, R.W., Dopita, M. A., Dekel, A., Hathi, N. P.,
    Straughn, A., \& Rutkowski, M. 2013, MNRAS, 429, L40

\bibitem[Lau \& Bertin(1978)]{lau78} Lau, Y.Y., \& Bertin, G. 1978, ApJ, 226, 508

\bibitem[Law et al.(2012)]{law12} Law, D.R., Shapley, A.E., Steidel, C.C., Reddy,
N.A., Christensen, C.R., \& Erb, D.K. 2012, Nature, 487, 338

\bibitem[Lee et al.(2013)]{lee}Lee, B. et al. 2013, ApJ, 774, 47

\bibitem[Lotz et al.(2006)]{lotz06} Lotz, J. M., Madau, P., Giavalisco, M.,
    Primack, J., \& Ferguson, H. 2006, ApJ, 636, 592

\bibitem[McLure et al.(2013)]{mclure13} McLure, R. J., Pearce, H. J.,
    Dunlop, J. S., et al. 2013, MNRAS, 428, 1088

\bibitem[Miyauchi-Isobe, Maehara, \& Nakajima(2010)]{kiso-all}
    Miyauchi-Isobe, N., Maehara, H., \& Nakajima, K. 2010,
    Pub.Nat.Astro.Ob.Japan, 13, 9

\bibitem[Oesch et al.(2007)]{oesch} Oesch, P.A. et al. 2007, ApJ, 671,  1212

\bibitem[Oesch et al.(2010)]{oesch10} Oesch, P.A. et al. 2010, ApJL, 709, L16

\bibitem[Puech et al.(2012)]{puech12} Puech, M., Hammer, F., Hopkins, P. F.,
    Athanassoula, E., Flores, H., Rodrigues, M., Wang, J. L., \& Yang, Y. B.
    2012, ApJ, 753, 128

\bibitem[Queyrel et al.(2012)]{queyrel12} Queyrel, J., Contini, T.,
Kissler-Patig, M., et al. 2012, A\&A, 539, A93

\bibitem[Rafelski et al.(2009)]{rafelski09} Rafelski, M., Wolfe, A.M., Cooke,
    J., Chen, H.-W., Armandroff, T.E., \& Wirth, G.D. 2009, ApJ, 703, 2033

\bibitem[Reddy et al.(2006)]{reddy}Reddy, N. et al. 2006, ApJ, 644, 792

\bibitem[Rix et al.(2004)]{rix04}  Rix, H.-W., Barden, M., Beckwith, S.V.W.,
Bell, E.F. et al. 2004, ApJS, 152, 163

\bibitem[Scoville et al.(2007)]{scoville}Scoville, N. et al. 2007. ApJS, 172, 1

\bibitem[Shapiro et al.(2010)]{shap10} Shapiro, K.L., Genzel, R., \&
    F\"orster Schreiber, N.M. 2010, MNRAS, 403, L36

\bibitem[Shapley(2011)]{shapley11} Shapley, A.E. 2011, ARA\&A, 49, 525

\bibitem[Sheth et al.(2012)]{sheth12} Sheth, K., Melbourne, J.,
    Elmegreen, D.M., Elmegreen, B.G., Athanassoula, E., Abraham, R.G., \& Weiner, B.J.
    2012, ApJ, 758, 136

\bibitem[Spergel et al.(2003)]{spergel} Spergel, D.N., et al. 2003, ApJS, 148, 175

\bibitem[Tacconi et al.(2010)]{tacconi10} Tacconi, L. J., et al. 2010, Nature, 463, 781

\bibitem[Tacconi et al.(2013)]{tacconi13} Tacconi, L. J., Neri, R., Genzel,
    R., et al. 2013, ApJ, 768, 74

\bibitem[van den Bergh(1960)]{vdb60} van den Bergh, S. 1960, ApJ, 131, 215

\bibitem[Wisnioski et al.(2011)]{wisnioski11} Wisnioski, E., Glazebrook,
    K., Blake, C., et al. 2011, MNRAS, 417, 2601

\bibitem[Wolf et al.(2003)]{wolf03} Wolf, C., Meisenheimer, K., Rix, H.-W.,
    Borch, A., Dye, S., \& Kleinheinrich, M. 2003, A\&A, 401, 73

\bibitem[Wright et al.(2009)]{wright09}Wright,  S. et al. 2009, ApJ, 699, 421

\end{thebibliography}
\end{document}